\documentclass[aps,pra,showpacs,twocolumn,groupedaddress,lengthcheck]{revtex4}
\usepackage{amsmath}
\usepackage{amssymb}
\usepackage{graphics}
\usepackage[dvips]{graphicx}
\usepackage{graphicx}
\usepackage{color}

\newcommand \beq{\begin{eqnarray}}
\newcommand \eeq{\end{eqnarray}}
\def\simge{\mathrel{%
       \rlap{\raise 0.511ex \hbox{$>$}}{\lower 0.511ex \hbox{$\sim$}}}}
\def\simle{\mathrel{
       \rlap{\raise 0.511ex \hbox{$<$}}{\lower 0.511ex \hbox{$\sim$}}}}

\newcommand \ga{\raisebox{-.5ex}{$\stackrel{>}{\sim}$}}

\begin{document}
\title{Normal mass density of a superfluid Fermi gas at unitarity }
\author{Gordon Baym$^{a,b}$ and C.\ J.\ Pethick$^{b,c}$}
\affiliation{\mbox{$^a$Department of Physics, University of Illinois, 1110
  W. Green Street, Urbana, IL 61801-3080} \\
\mbox{$^b$The Niels Bohr International Academy, The Niels Bohr Institute,}\\\mbox{ University of Copenhagen, Blegdamsvej 17, DK-2100 Copenhagen \O,
 Denmark}\\
\mbox{$^c$NORDITA, KTH Royal Institute of Technology and Stockholm University,}\\
\mbox{
Roslagstullsbacken 23, SE-10691 Stockholm, Sweden }\\
}

\date{\today}

\begin{abstract}
We calculate the normal mass density of a paired Fermi gas at unitarity.  The dominant contribution near the superfluid transition is from fermionic quasiparticle excitations, and is thus sensitive to the pairing gap.   A comparison with the recent experiment of Sidorenkov et al. [Nature {\bf 498}, 78 (2013)] suggests that the superfluid gap near the transition temperature is larger than the BCS value, but the data do not permit a quantitative inference of the gap.  Calculations of the quenched moment of inertia of a BCS superfluid in a harmonic trap are in reasonable agreement with the earlier experiment of Riedl et al. [New J. Physics {\bf 13}, 035003 (2011)].
   
\end{abstract}

\maketitle

\section{Introduction}

  The superfluid mass density, $\rho_s$, of a superfluid Fermi gas at unitarity is a sensitive probe of the behavior of the quasiparticle structure of the system near the transition temperature \cite{rhos}.  Indeed, Sidorenkov et al.~have recently provided such a measurement of $\rho_s$ in a paired Fermi gas of $^6$Li atoms at unitarity \cite{grimm}, which complements earlier measurements of the suppression of the moment of inertia in a rotating  paired gas of $^6$Li \cite{grimminertia}.  This latter suppression of the moment of inertia arises from the decrease of the normal mass density, $\rho_n = \rho - \rho_s$ (where $\rho$ is the total mass density), with decreasing temperature.   Our goal in this paper is to provide a framework for understanding the normal mass density of Fermi gases at unitarity.   

   The Leggett model \cite{tony} of pairing at unitarity in terms of the BCS wave function indicates that the physics is much closer to BCS, with a positive chemical potential, than it is to BEC.  Thus a reasonable starting point in understanding the normal mass density, which we adopt in this paper, is to model the superfluid mass density on the BCS-like structure of the Fermi superfluid. 

  The normal mass density of a superfluid is composed of contributions from the excitations present at
finite temperature. 
 In a Fermi superfluid the excitations include fermionic quasiparticles, as well as the bosonic collective modes -- first and second sound.  The dominant contribution to the normal mass density of a paired Fermi superfluid is that of the fermionic quasiparticle excitations.  As we show, both first and second sound contribute negligibly to the normal mass density near the transition temperature.   Second sound, although it has a significantly smaller velocity than first sound, becomes highly damped at finite wavevector owing to thermal conduction and viscosity, which limits its contribution to the normal mass density. 

  The normal mass density is given exactly in terms of the Fourier transformation of the transverse current-current correlation commutator, $\langle \left[ j_\perp(rt), j_\perp(r't')\right]\rangle \equiv \Upsilon_\perp(r-r',t-t')$, by \cite{jj} 
   \beq
  \rho_n = \lim_{k\to0}\,m^2\int \frac{d\omega}{2\pi}
   \frac{\Upsilon_\perp(k,\omega)}{\omega}.
\label{rhon}
\eeq
For independent long-lived quasiparticles this expression yields $\rho_n = \sum_i  \rho_n^{(i)}$, where 
following Landau's argument \cite{landau1941}, each branch of excitation, $i$, makes a contribution to the normal mass density,
\beq
  \rho_n^{(i)}= -\int \frac{d^3k}{(2\pi)^3} \frac{k^2}{3}\frac{\partial f_i(k)}{\partial \epsilon_i(k)},
  \label{landau}
\eeq
where $\epsilon_i(k)$ is the energy of an excitation of momentum $k$ in branch $i$, and $f_i(k)$ is the usual Bose or Fermi distribution function.  (In this paper we take $\hbar = 1 = k_B$ throughout.)  Since first and second sound are collective modes of the fermionic excitations,
this expression somewhat overcounts the weight of the fermionic quasiparticle excitations, an effect which can be neglected to a first approximation.  Numerical calculations of $\rho_n$ in terms of correlation functions for paired Fermi systems are given in Ref.~\cite{strinati}, while Ref.~\cite{salasnich} calculates the normal mass density using Eq.~(\ref{landau}) with
particular approximate expressions for the first sound and quasiparticle spectra.

\section{Fermionic quasiparticle contributions to the normal mass density}

    We consider now a homogeneous Fermi gas  with equal populations of two components of the same mass.
To a first approximation at unitarity,  the fermionic quasiparticle excitation energies have the form:
\beq
  E_p = \left\{[\epsilon_p - \mu(T)]^2   +\Delta(T)^2  \right\}^{1/2},
  \label{qpart0}
\eeq
where $\epsilon_p$ is the quasiparticle energy in the absence of pairing, $\mu(T)$ is the chemical potential, and  $\Delta(T)$ is the pairing gap at temperature $T$.  For a unitary Fermi gas, the chemical potential has the form $\mu=\xi T_F$, where $\xi$ is a function only of $T/T_F$, with $T_F$ the Fermi temperature, $p^2_F /2m^*$ (with the Fermi momentum given by $p_F=(3\pi^2n)^{1/3}$,  $n$ the total fermion density, and $m^*$ the fermion effective mass).
At $T=0$, $\xi(0) \simeq 0.376(4)$ \cite{mz,jc}.
The $\epsilon_p$ are given essentially by kinetic energies, $\epsilon_p^0 = p^2/2m^*$, plus self-energy shifts, $U$.  Writing $\mu - U = \epsilon_{\rm min}(T)$, we have then
\beq
  E_p = \left\{[\epsilon_p - \epsilon_{\rm min}(T)]^2   +\Delta(T)^2  \right\}^{1/2},
  \label{qpart}
\eeq

At zero temperature, $\epsilon_{\rm min}$ falls from the Fermi energy, $\epsilon_F (= p^2_F /2m^*)$, in the
weak coupling BCS limit  to $\sim$ 0.8 - 0.9 $\epsilon_F$ at unitarity \cite{CR05,mag}.  In contrast the  BCS wave function at unitarity yields $\epsilon_{\rm min}(T=0) \sim 0.6 \epsilon_F$.  However, as we find below, the fermionic quasiparticle contribution to the normal mass density is relatively insensitive to the detailed value of  $\epsilon_{\rm min}(T)$.    In addition, Monte Carlo calculations in Ref.~\cite{mag} show that the effective mass is
very little shifted from the bare mass, indicating that  the Landau parameter $F_1^S =3(m^*/m-1) $, which determines the strength of current--current interactions between quasiparticles,  is close to zero.  Below we take $m^*=m$.

   The fermionic quasiparticles, with energy $E_p$,  contribute
\beq
 \rho_n^{(qp)}&=& -2\int \frac{d^3p}{(2\pi)^3} \frac{p^2}{3}\frac{\partial}{\partial E_p}\frac{1}{e^{\beta E_p} +1}\nonumber \\
 &=&  \frac{\beta}{6}\int \frac{d^3p}{(2\pi)^3} p^2  {\rm sech}^2(\beta E_p/2),
 \eeq
 to the normal mass density, where $\beta = 1/T$.
Since this contribution is dominant, we drop the ``qp" label in this section.  We assume, in evaluating $\rho_n$ using Eq.~(\ref{qpart}), that the gap has the mean-field form near $T_c$,
\beq
  \Delta(T) = \Delta_c\sqrt{1-T/T_c},
  \label{dc}
\eeq 
and we regard $\Delta_c$ as a fittable parameter.  

  In the immediate neighborhood of $T_c$, the superfluid density with a mean-field gap, Eq.~(\ref{dc}), vanishes linearly as $T$ approaches $T_c$.   In BCS \cite{LP},
\beq
  \frac{\rho_s}{\rho} = \frac{7\zeta(3)}{4\pi^2}\left(\frac{\Delta_c}{T_c}\right)^2\left(1-\frac{T}{T_c}\right),
\eeq
and since, 
\beq
\Delta_c  =\sqrt{8\pi^2/7\zeta(3)}T_c= 1.74\Delta(T=0) = 3.06 T_c,
\label{BCSgap}
\eeq 
the slope is exactly two.   One should not expect this result to necessarily hold at unitarity, given that the measured zero temperature gap is $\sim$ 0.44 $T_F$ \cite{sch}, and $\sim$ 0.45 $T_F$ in Monte Carlo at unitarity \cite{CR}, corresponding to \cite{jin} $\sim 2.6 T_c$, compared with 1.76 $T_c$ in BCS.   See also  Refs.~\cite{mag,wz}.   Were the gap uniformly increased from BCS by a temperature independent constant, one would expect the approach of $\rho_n/\rho$ to zero as $T\to T_c$ to have a greater slope than in BCS; however, as one sees in Fig.~1 the slope is quite close to two. 

      Using Eq.~(\ref{qpart}) for $E_p$, we have
\beq
  \frac{\rho_n}{\rho} = \frac{ \int dp\, p^4 {\rm sech}^2(\beta E_p/2 )}
 {\int dp \,p^4   {\rm sech}^2 (\beta [p^2/2m-\epsilon_{\rm min}(T)]/2)};
\eeq
for $\Delta = 0$, the ratio explicitly equals unity .

\begin{figure}[h]
\begin{center}
\includegraphics[width=8.5cm,height=7cm]{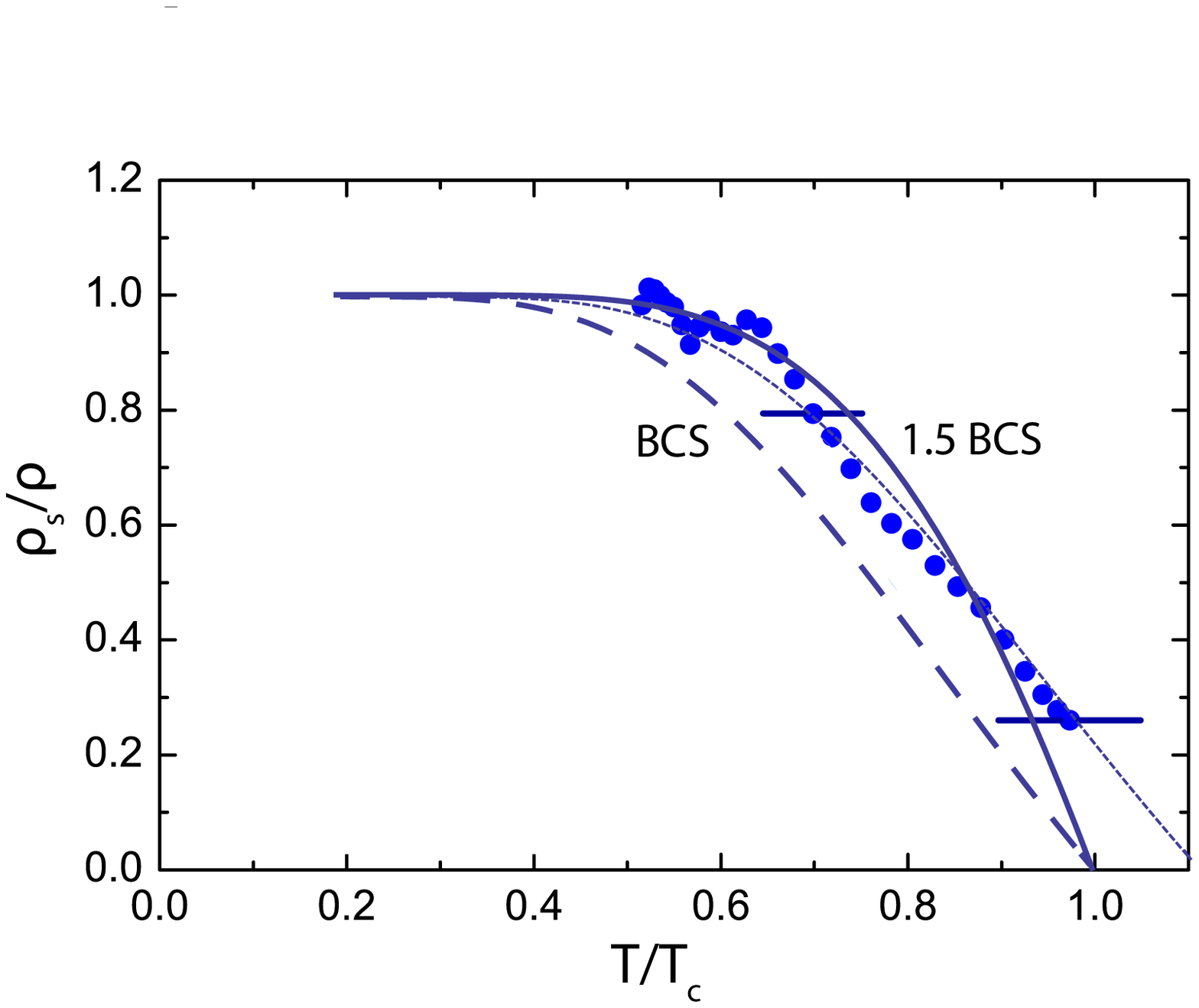}
\caption{(Color online) Calculated superfluid mass density compared with the experimental measurement of Ref.~\cite{grimm}.  The dashed curve labeled BCS uses the BCS gap structure near $T_c$, Eqs.~(\ref{dc},\ref{BCSgap}), while the solid curve uses a gap a factor of 1.5 larger.  The thin dotted curve is the BCS curve uniformly stretched in temperature to pass optimally through the data.
}
\label{data}
\end{center}
\end{figure}  
  
   Figure~1 shows the results of numerical integration of this expression for $\rho_s/\rho = 1-\rho_n/\rho$, with parameters $T_c/T_F = 0.167$ \cite{mz},  and $\Delta_c/T_c = 3.06$ as in BCS, Eq.~(\ref{BCSgap}), as well as 
$\Delta_c/T_c = 4.6$. a factor 2.6/1.74  larger than the BCS result.   We take $\epsilon_{\rm min}(T) = \epsilon_F$ here.
The data points are from Ref.~\cite{grimm}.  As we see, the BCS result begins to fall from unity at $T/T_c \simeq 0.4$;  the start of the experimental falloff at $T/T_c \,\ga\, 0.6$ is consistent with a larger $\Delta_c/T_c$ than in BCS, but can also be attributed to the uncertainty in the absolute temperature scale.  Note that the slope at temperatures near $T_c$ is closer to that of the BCS prediction.   We also show in Fig.~1 as a thin dotted line, the BCS result stretched by a scale change of $T$, which would be equivalent to scaling the temperatures corresponding to the data points downward by about 10\%, an amount consistent with the horizontal error bars shown in the data.  This latter agreement can be interpreted as the data showing a slope two, as in BCS.

\subsection{Suppression of the moment of inertia}

   We next apply the present analysis to the measurement in Ref.~\cite{grimminertia} in a harmonic trap of the suppression of the moment of inertia in the presence of superfluidity.  (For background theoretical discussion of slowly rotating trapped paired fermion systems, see Ref.~\cite{urban} and references therein.) In a trap the local chemical potential obeys,
 \beq
  \mu({\bf r},T) - \frac{m}{2} (\omega_z^2 z^2 + \omega_\perp r_\perp^2) =  \mu(T),
    \label{p}
 \eeq
where $\mu(T)\equiv \mu({\bf r}=0,T)$ is the chemical potential of the system.  We assume, for effects near $T_c$, that $\mu({\bf r}) = \xi_c T_F({\bf r})$, where $\xi_c \equiv \xi(T_c/T_F)$.    With Eq.~(\ref{p}), the total particle number is given by
\beq
   N = \int d^3r\,n(r)=\frac{1}{24}p_F(0)^3R_\perp^2Z,
\eeq
where $n(r)  = p_F({\bf r})^3/3\pi^2$, the transverse radius $R_\perp = p_F(0)\sqrt{\xi_c}/m\omega_\perp$, and $Z= p_F(0)\sqrt{\xi_c}/m\omega_z$.  
Similarly the normal state or classical moment of inertia about the z axis is
\beq
  I_{cl} =  \int d^3r\, mr_\perp^2 n({\bf r})=\frac{m}{4}NR_\perp^2.
\eeq  

   With increasing temperature the superfluid region in the trap shrinks, and  the moment of inertia of the superfluid 
component is given by
\beq
  I_s(T) =  \int d^3r\, r_\perp^2\rho_s({\bf r}).
  \eeq  
Rather than using the full expression for the local $\rho_s({\bf r})$, we assume that  $\rho_s({\bf r})/\rho(r)$ near $T_c$ falls to zero as $1-T^2/T_c({\bf r})^2$ (cf. Fig.~1), where $T_c(r) = 0.167 \,T_F(r)$ is the local 
superfluid transition temperature.  Then
\beq
  \frac{I_s}{I_{cl}} = 
   \frac{256}{\pi}\int_0^{x_c} dx \,x^4 (1-x^2)^{3/2} \left[ 1-\left(\frac{1-x_c^2}{1-x^2}\right)^2   \right], \nonumber\\
  \label{isicl}
\eeq
where $x_c = \left(1-T/T_c(0)\right)^{1/2}$ and $T_c(0)$ is the transition temperature in the center of the trap.    The normal moment of inertia ratio is then
\beq
  \frac{I_n(T)}{I_{cl}} = 1 -   \frac{I_s(T)}{I_{cl}} =   \frac{1028}{35\pi}\left(1-\frac{T}{T_c(0)}\right)^{7/2} .
\eeq  
The relative suppression of the  moment of inertia is independent of the value of $\xi_c$.

   The experiment of Riedl et al.~\cite{grimminertia} measures the total angular momentum, $L=I_n\Omega$, of paired $^6$Li at unitarity,  driven at angular velocity $\Omega_{trap}$ and rotating at angular frequency $\Omega$.   The method is to measure the precession frequency of a radial quadrupole excitation, $\Omega_{prec} = L/2I_{cl}$ and from this frequency to infer the precession parameter,
\beq
{\cal P}(T)  =  \frac{I_n(T)}{I_{cl}} \frac{\Omega}{\Omega_{trap}}.
\eeq
The data for $\cal P$ is shown in the inset in Fig. 2 as a function of the ratio of an inferred temperature $\cal T$ to $T_{F,trap}$, here the Fermi temperature of a {\it free} gas in the center of the trap; $T_{F,trap}$ is smaller than the Fermi temperature at the center of the trap by a factor $\sim 0.8$.  The transition temperature is $T_c \sim
0.21T_{F,trap}$.

    Figure~2 compares the calculated $I_n(T)/I_{cl}$ with the data of Ref.~\cite{grimminertia}.    There are two uncertainties in this comparison; the first is the ratio $\Omega/\Omega_{trap}$, which equals unity in a steady state and is a few percent less in practice, and the second is the measured temperature $\cal T$, which lies somewhat below the true temperature $T$.   In Fig.~2 we have taken $T_c = 0.21 T_{F,trap}$,  and have scaled the data for $\cal P$ vertically upwards by a few percent, consistent with the magnitude of $1-\Omega/\Omega_{trap}$.   Given these uncertainties the agreement of theory with the scaled data is satisfactory.   The data also shows the slow approach in a trap, $\sim (1-T/T_c(0))^{7/2}$, of the normal moment of inertia to $I_{cl}$, Eq.~(\ref{isicl}).  The overall agreement points to the desirability of further measurements of the moment of inertia with a carefully calibrated temperature scale.

\begin{figure}[h]
\begin{center}
\includegraphics[width=8.5cm,height=7cm]{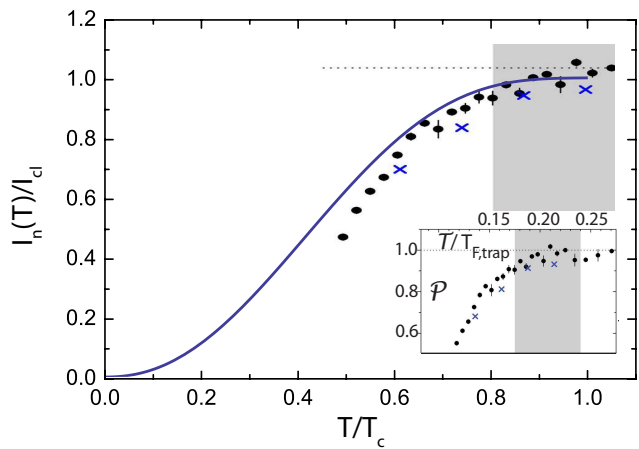}
\caption{(Color online) Calculated moment of inertia in units of the normal state moment of inertia in a trap. 
The data along the theoretical curve has been scaled upward by a few percent from the original data for the precession parameter ${\cal P}$ of Ref.~\cite{grimminertia}; see text for details.  The inset shows the original data. }
\label{mom}
\end{center}
\end{figure}  

\section{Contributions of collective modes}

We now estimate the contributions of first and second sound in a homogeneous system to the normal mass density.   Collective modes in a trap are discussed experimentally and theoretically in Refs.~\cite{collective,hou}.

\subsection{First sound}

The standard Landau result for the contribution of first sound of velocity $s$ to the normal mass density is
\beq
  \rho_n^{(1)} = \frac {2\pi^2}{45}\frac{T^4}{s^5}.
\eeq
The first sound velocity is
 given generally  by 
$ms^2 = \partial^2 F/\partial n^2$,
where $F$ is the free energy density, and $m$ is the bare atomic mass.   At zero temperature
we may write $F = \xi(0) n \epsilon_F$.     Thus
\beq
   s = \sqrt{\xi(0)/3}\,\,v_F = 0.35\,v_F,
\eeq
with $v_F=p_F/m$ being the Fermi velocity, a result
consistent with the measurement reported in Ref.~\cite{grimm}.
The relative first sound contribution to the normal mass density then has the form 
\beq
  \frac{\rho_n^{(1)}}{\rho} \simeq \frac{3^{2/3}\pi^4}{40\xi(0)^{5/2} } \left(\frac{T_c}{T_F}\right)^4 \left(\frac{T}{T_c}\right)^4 \simeq 0.11 \left(\frac{T}{T_c}\right)^4;
\eeq
Here we use $T_c \simeq 0.167 \,T_F$ \cite{mz}.   The first sound contribution to the normal mass density is relatively small near the transition temperature.  This result is in reasonable agreement with numerical evaluation of $\rho_n^{(1)}/\rho$ by Andrenacci et al.~\cite{strinati} for $T/T_c\le 0.5$.

\subsection{Second sound}

   Second sound is, to a first approximation, an excitation of the temperature at fixed particle chemical potential.   In Appendix A, we review the derivation of the velocity and damping of second sound.  As seen there the
hydrodynamic dispersion relation of second sound, neglecting the coupling of temperature and chemical potential fluctuations, is
 \beq
  \omega^2 = s_2^2 k^2 - 2i\omega \Gamma,
  \label{damp0}
 \eeq
with the second sound velocity, $s_2$, given by
 \beq
   s_2^2 = \frac{\rho_s}{\rho_n}\frac{S^2}{\rho\, \partial S/\partial T},
 \eeq 
and the damping rate, $\Gamma$, by
\beq
\Gamma=\frac{k^2}{2}\left[\frac{K}{C}+  \frac{\rho_s}{\rho_n}\left(  \frac43 \frac{\eta}{\rho} +\frac{\zeta_2}{\rho}-2\zeta_1+\rho\zeta_3   \right)        \right].
\label{Gamma}
\eeq
Here $K$ is the thermal conductivity, $C=T\partial S/\partial T$ the heat capacity per unit volume, $\eta$  the shear viscosity and $\zeta_2$ the bulk viscosity of the normal fluid. The quantity $\zeta_3$ is a bulk viscosity associated with the motion of the superfluid, and $\zeta_1$  a bulk viscosity associated with motion of both fluids.  For a Fermi superfluid the bulk viscosities have not been investigated.  However, Putterman has shown that the contribution of the bulk viscosities to the damping of second sound vanishes if the non-equilibrium  state can be described in terms of a single macroscopic variable (in addition to those occuring in the hydrodynamic equations) that relaxes to a value dependent only on the hydrodynamic variables \cite{putterman}.  This argument suggests that damping by bulk viscosities is small, and we neglect it.   The microscopic mechanisms for bulk viscosities in Fermi superfluids should be investigated in detail in the future.  Near $T_c$ the contribution to the damping from shear viscosity is suppressed compared with that from thermal conduction by a factor $\rho_s/\rho_n$, and therefore to estimate the damping rate we consider only the contribution from thermal conduction:\beq
   \Gamma \approx  \frac12 D_T k^2 ,
\eeq
where $D_T=K/C$ is the thermal diffusivity.  This provides a lower bound on the damping.
 
   Generally, $s_2 $ is small compared with $s$, tending to zero as $T\to T_c$; thus naively one would expect a very large contribution from second sound to the normal mass density.   However, the contribution is limited by the damping of second sound modes at finite wavevectors.  The thermal diffusivity is of order $v_F \ell$, where $\ell$ is the mean free path.   Thus from Eq.~(\ref{damp0}), the imaginary part of the second sound dispersion equals the real part at wavevector
\beq  
  k_{\rm max}= s_2/D_T.
 \eeq
Including viscous damping would only decrease $k_{\rm max}$.

  Such damping implies that second sound is a well-defined collective mode only for $k < k_{\rm max}$, and therefore in the integral for the contribution of second sound to $\rho_n$ one should exclude modes with larger wave numbers.  We thus  estimate,
\beq 
    \rho_n^{(2)} \simeq -\int_0^{k_{\rm max}}\frac{d^3k}{(2\pi)^3} \frac{k^2}{3s_2}\frac{\partial}{\partial k}\frac{1}{e^{\beta s_2 k}-1}.
    \label{rhosecond}
\eeq
 In addition, the frequency of second sound at $k_{\rm max}$ is small compared with the critical temperature; we have
\beq
 \frac{s_2 k_{\rm max}}{T_c} \sim \left(\frac{s_2}{v_F}\right)^2\frac{T_F/T_c}{\ell p_F},
 \eeq
and since $p_F \ell \gg 1$ and $s_2^2\ll v_F^2=(p_F/m)^2$, the right side is smaller than unity.  Thus to a first approximation we expand the exponential in the distribution function in Eq.~(\ref{rhosecond}) for temperature near $T_c$, and find
 \beq
   \frac{\rho_n^{(2)}}{\rho} \sim \frac{T}{6ms_2^2}  \left(\frac{k_{\rm max}}{p_F}\right)^3 \sim \frac{T}{m v_F^2}\frac{s_2}{v_F}\left(\frac{1}{p_F\ell}\right)^3.
  \label{rhon2}
    \eeq  
In a unitary Fermi gas at $T_c$ the mean free path is of order the particle spacing, but as the temperature is lowered, the mean free path becomes longer due to increased Pauli blocking and the reduction of the number of thermal excitations from which to scatter.   Estimates for the viscous mean free path suggest 
that the minimum value of $\ell$ is about $3/k_F$ \cite{riedl}.  Since $T_c/mv_F^2=T_c/2T_F\approx 0.08$ and $s_2/v_F$ is typically around 0.1 or less, Eq.\ (\ref{rhon2}) shows that second sound  makes a negligible contribution to the normal mass density near $T_c$.  

\section{Conclusion}

     We have given here a framework for determining the normal mass density of a paired Fermi superfluid near the transition temperature.  To a first approximation, the data are described in terms of  BCS-like excitations of the paired superfluid and, as we have shown. the experiments are a sensitive probe of the pairing gap,.  However, to go beyond the present simple calculation, considerably more physics needs to be included.   The list includes an improved understanding of the quasiparticle structure beyond Eq.~(\ref{qpart}), together with quasiparticle damping effects. One also needs to include Fermi liquid effects; for example, when Fermi liquid effects are included within the BCS approximation the slope of $\rho_n$ near $T_c$ is reduced by a factor $m/m*=1/(1+F^s_1/3)$ \cite{leggett}. 
     
       Further experimental studies of second sound, and its damping in particular, would provide valuable information about the transport properties of the system.   These need to be of high precision because of the difficulties associated with deducing bulk properties of a homogeneous system from data on atoms in traps, where matter is inhomogeneous.  In measurements for trapped atoms it is difficult to determine $T_c$ because just below $T_c$ matter is superfluid in only a small volume at the center of the trap. 
     
      It is also necessary to understand better the interplay between single quasiparticle  and collective effects, such as those that lead to first and second sound.   One such effect is the coupling between first and second sound due to the nonzero thermal expansion of the system, an effect not taken into account above, but which has been found to be significant in elongated traps \cite{hou}.  The detailed microscopic mechanisms that give rise to the bulk viscosities should be investigated;  among processes that play a role here is the one that has been studied extensively in nonequilibrium superconductors, where the condensate and the BCS-like excitations may to a first approximation be regarded as having different chemical potentials \cite{PS}.

\section*{Acknowledgements}

We thank G. Bruun, A. Gezerlis, T.-L. Ho, W. Zwerger, M. Zwierlein, and R. Grimm for very helpful discussions. Author GB is grateful for the hospitality of the Niels Bohr International Academy and the BEC Center in Trento, where parts of
this work were carried out.   This research was supported in part by NSF Grants PHY09-69790 and PHY13-05891.

\appendix
\section{Second sound}
   
    In this Appendix we review the derivation of the velocity and damping of second sound.   One begins with the conservation of momentum, which takes the linearized form,
 \beq
 \rho_s \frac{\partial \vec v_s}{\partial t} +  \rho_n\frac{\partial \vec v_n}{\partial t} + 
 \nabla P& =  \eta \nabla^2 \vec v_n + \left(\zeta_2 + \eta/3) \nabla(\nabla\cdot  \vec v_n \right) \nonumber \\
 & +\zeta_1 \rho_s  \nabla \left( \nabla\cdot(\vec v_s-\vec v_n)\right),
 \eeq
 where $\vec v_s$ is the superfluid velocity, $\vec v_n$ the normal fluid velocity, and $P$ is the pressure; and the superfluid acceleration equation,
 \beq 
   m\frac{\partial \vec v_s}{\partial t} +  \nabla \mu = \zeta_3\rho_s \nabla \left( \nabla\cdot(\vec v_s-\vec v_n)\right)
   +\zeta_4 \nabla(\nabla\cdot\vec v_n). \nonumber\\
 \eeq
Here $\eta$ is the first viscosity, and the $\zeta_i$ are the four second viscosities; the Onsager reciprocity relations imply
that $\zeta_4=\zeta_1$.
We derive here only the contribution of thermal conductivity to the damping, which is expected to give the dominant contribution to, as well as a lower bound on, the damping rate of second sound.   The full result, including the viscosities is given in Eq.~(\ref{Gamma}).
The two acceleration equations, together with the Gibbs-Duhem relation,
$dP = nd\mu + SdT$,  imply
 \beq
 \frac{\partial \vec v_n}{\partial t} +\frac{\rho_s}{\rho_n\rho} S\nabla T +  \frac{1}{\rho}\nabla P = 0,
  \label{vn}
 \eeq
 In addition, the linearized equation for entropy conservation is 
\beq
\frac{\partial S}{\partial t} + S\nabla\cdot\vec v_n - \frac{K}{T}\nabla^2 T = 0,
\label{s}
\eeq
where $K$ is the thermal conductivity.   Using Eqs.~(\ref{vn}) at fixed $P$, and (\ref{s}) to eliminate $\vec v_n$ 
we find the equation of second sound propagation,
\beq
\frac{\partial^2 \delta S}{\partial t^2} - \frac{\rho_s}{\rho_n}\frac{S^2}{\rho} \nabla^2 \delta T - \frac{K}{T} \nabla^2 \frac{\partial}{\partial t} \delta T = 0,
 \eeq
 where for clarity we write $\delta$ here to indicate the first order terms.  Writing $\delta S = (dS/dT)\delta T$, we find the
 hydrodynamic dispersion relation of second sound,
 \beq
  \omega^2 = s_2^2 k^2 - i\omega k^2 \frac{K}{T\partial S/\partial T},
  \label{damp}
 \eeq
 with the second sound velocity given by
 \beq
   s_2^2 = \frac{\rho_s}{\rho_n}\frac{S^2}{\rho\, \partial S/\partial T}.
 \eeq


\begin{thebibliography}{20}
\bibitem{rhos}  The superfluid mass density measures the momentum density, $\vec g$, carried by a superfluid flow of velocity $\vec v_s = (\hbar/2m)\nabla \phi$, viz., $\vec g = \rho_s \vec v_s$. where $\phi$ is the phase of the superfluid order parameter (the condensate wavefunction).
\end{thebibliography}
\end{document}